\documentclass[conference]{IEEEtran}
\IEEEoverridecommandlockouts
\usepackage{cite}
\usepackage{caption}
\usepackage{amsmath,amssymb,amsfonts}
\usepackage{algorithmic}
\usepackage{graphicx}
\usepackage{caption}
\usepackage{subcaption}
\usepackage{textcomp}
\usepackage{xcolor}
\usepackage{url}
\usepackage{multirow} 

\usepackage[acronyms,nonumberlist,nopostdot,nomain,nogroupskip,acronymlists={hidden}]{glossaries}
\newglossary[algh]{hidden}{acrh}{acnh}{Hidden Acronyms}
\newacronym{3gpp}{3GPP}{3rd Generation Partnership Project}
\newacronym{4g}{4G}{4th generation}
\newacronym{5g}{5G}{5th generation}
\newacronym{6g}{6G}{6th generation}
\newacronym{5gc}{5GC}{5G Core}
\newacronym{adc}{ADC}{Analog to Digital Converter}
\newacronym{aerpaw}{AERPAW}{Aerial Experimentation and Research Platform for Advanced Wireless}
\newacronym{ai}{AI}{Artificial Intelligence}
\newacronym{af}{AF}{Application Function}
\newacronym{aimd}{AIMD}{Additive Increase Multiplicative Decrease}
\newacronym{am}{AM}{Acknowledged Mode}
\newacronym{amc}{AMC}{Adaptive Modulation and Coding}
\newacronym{amf}{AMF}{Access and Mobility Management Function}
\newacronym{aops}{AOPS}{Adaptive Order Prediction Scheduling}
\newacronym{api}{API}{Application Programming Interface}
\newacronym{apn}{APN}{Access Point Name}
\newacronym{ap}{AP}{Application Protocol}
\newacronym{aqm}{AQM}{Active Queue Management}
\newacronym{ar}{AR}{Augmented Reality}
\newacronym{ausf}{AUSF}{Authentication Server Function}
\newacronym{avc}{AVC}{Advanced Video Coding}
\newacronym{awgn}{AGWN}{Additive White Gaussian Noise}
\newacronym{balia}{BALIA}{Balanced Link Adaptation Algorithm}
\newacronym{bbu}{BBU}{Base Band Unit}
\newacronym{bdp}{BDP}{Bandwidth-Delay Product}
\newacronym{ber}{BER}{Bit Error Rate}
\newacronym{bf}{BF}{Beamforming}
\newacronym{bler}{BLER}{Block Error Rate}
\newacronym{brr}{BRR}{Bayesian Ridge Regressor}
\newacronym{bs}{BS}{Base Station}
\newacronym{bsr}{BSR}{Buffer Status Report}
\newacronym{bss}{BSS}{Business Support System}
\newacronym{ca}{CA}{Carrier Aggregation}
\newacronym{caas}{CaaS}{Connectivity-as-a-Service}
\newacronym{cb}{CB}{Code Block}
\newacronym{cc}{CC}{Congestion Control}
\newacronym{ccid}{CCID}{Congestion Control ID}
\newacronym{cco}{CC}{Carrier Component}
\newacronym{cdd}{CDD}{Cyclic Delay Diversity}
\newacronym{cdf}{CDF}{Cumulative Distribution Function}
\newacronym{cdn}{CDN}{Content Distribution Network}
\newacronym{cn}{CN}{Core Network}
\newacronym{codel}{CoDel}{Controlled Delay Management}
\newacronym{comac}{COMAC}{Converged Multi-Access and Core}
\newacronym{cord}{CORD}{Central Office Re-architected as a Datacenter}
\newacronym{cornet}{CORNET}{COgnitive Radio NETwork}
\newacronym{cosmos}{COSMOS}{Cloud Enhanced Open Software Defined Mobile Wireless Testbed for City-Scale Deployment}
\newacronym{cots}{COTS}{Commercial Off-the-Shelf}
\newacronym{cp}{CP}{Control Plane}
\newacronym{cyp}{CPR}{Cyclic Prefix}
\newacronym{up}{UP}{User Plane}
\newacronym{cpu}{CPU}{Central Processing Unit}
\newacronym{cqi}{CQI}{Channel Quality Information}
\newacronym{cr}{CR}{Cognitive Radio}
\newacronym{cran}{C-RAN}{Cloud \gls{ran}}
\newacronym{crs}{CRS}{Cell Reference Signal}
\newacronym{csi}{CSI}{Channel State Information}
\newacronym{cir}{CIR}{Channel Impulse Response}
\newacronym{csirs}{CSI-RS}{Channel State Information - Reference Signal}
\newacronym{cu}{CU}{Central Unit}
\newacronym{d2tcp}{D$^2$TCP}{Deadline-aware Data center TCP}
\newacronym{d3}{D$^3$}{Deadline-Driven Delivery}
\newacronym{dac}{DAC}{Digital to Analog Converter}
\newacronym{dag}{DAG}{Directed Acyclic Graph}
\newacronym{das}{DAS}{Distributed Antenna System}
\newacronym{dash}{DASH}{Dynamic Adaptive Streaming over HTTP}
\newacronym{dc}{DC}{Dual Connectivity}
\newacronym{dccp}{DCCP}{Datagram Congestion Control Protocol}
\newacronym{dce}{DCE}{Direct Code Execution}
\newacronym{dci}{DCI}{Downlink Control Information}
\newacronym{dctcp}{DCTCP}{Data Center TCP}
\newacronym{dl}{DL}{Downlink}
\newacronym{dmr}{DMR}{Deadline Miss Ratio}
\newacronym{dmrs}{DMRS}{DeModulation Reference Signal}
\newacronym{drlcc}{DRL-CC}{Deep Reinforcement Learning Congestion Control}
\newacronym{drs}{DRS}{Discovery Reference Signal}
\newacronym{du}{DU}{Distributed Unit}
\newacronym{e2e}{E2E}{end-to-end}
\newacronym{ecaas}{ECaaS}{Edge-Cloud-as-a-Service}
\newacronym{ecn}{ECN}{Explicit Congestion Notification}
\newacronym{edf}{EDF}{Earliest Deadline First}
\newacronym{embb}{eMBB}{Enhanced Mobile Broadband}
\newacronym{empower}{EMPOWER}{EMpowering transatlantic PlatfOrms for advanced WirEless Research}
\newacronym{enb}{eNB}{evolved Node Base}
\newacronym{endc}{EN-DC}{E-UTRAN-\gls{nr} \gls{dc}}
\newacronym{epc}{EPC}{Evolved Packet Core}
\newacronym{eps}{EPS}{Evolved Packet System}
\newacronym{es}{ES}{Edge Server}
\newacronym{etsi}{ETSI}{European Telecommunications Standards Institute}
\newacronym[firstplural=Estimated Times of Arrival (ETAs)]{eta}{ETA}{Estimated Time of Arrival}
\newacronym{eutran}{E-UTRAN}{Evolved Universal Terrestrial Access Network}
\newacronym{faas}{FaaS}{Function-as-a-Service}
\newacronym{fapi}{FAPI}{Functional Application Platform Interface}
\newacronym{fdd}{FDD}{Frequency Division Duplexing}
\newacronym{fdm}{FDM}{Frequency Division Multiplexing}
\newacronym{fdma}{FDMA}{Frequency Division Multiple Access}
\newacronym{fed4fire}{FED4FIRE+}{Federation 4 Future Internet Research and Experimentation Plus}
\newacronym{fir}{FIR}{Finite Impulse Response}
\newacronym{fit}{FIT}{Future \acrlong{iot}}
\newacronym{fpga}{FPGA}{Field Programmable Gate Array}
\newacronym{fr2}{FR2}{Frequency Range 2}
\newacronym{fs}{FS}{Fast Switching}
\newacronym{fscc}{FSCC}{Flow Sharing Congestion Control}
\newacronym{ftp}{FTP}{File Transfer Protocol}
\newacronym{fw}{FW}{Flow Window}
\newacronym{ge}{GE}{Gaussian Elimination}
\newacronym{gnb}{gNB}{5G base station}
\newacronym{gop}{GOP}{Group of Pictures}
\newacronym{gpr}{GPR}{Gaussian Process Regressor}
\newacronym{gpu}{GPU}{Graphics Processing Unit}
\newacronym{gtp}{GTP}{GPRS Tunneling Protocol}
\newacronym{gtpc}{GTP-C}{GPRS Tunnelling Protocol Control Plane}
\newacronym{gtpu}{GTP-U}{GPRS Tunnelling Protocol User Plane}
\newacronym{gtpv2c}{GTPv2-C}{\gls{gtp} v2 - Control}
\newacronym{gw}{GW}{Gateway}
\newacronym{harq}{HARQ}{Hybrid Automatic Repeat reQuest}
\newacronym{hetnet}{HetNet}{Heterogeneous Network}
\newacronym{hh}{HH}{Hard Handover}
\newacronym{hol}{HOL}{Head-of-Line}
\newacronym{hqf}{HQF}{Highest-quality-first}
\newacronym{hss}{HSS}{Home Subscription Server}
\newacronym{http}{HTTP}{HyperText Transfer Protocol}
\newacronym{ia}{IA}{Initial Access}
\newacronym{iab}{IAB}{Integrated Access and Backhaul}
\newacronym{ic}{IC}{Incident Command}
\newacronym{ietf}{IETF}{Internet Engineering Task Force}
\newacronym{imsi}{IMSI}{International Mobile Subscriber Identity}
\newacronym{imt}{IMT}{International Mobile Telecommunication}
\newacronym{iot}{IoT}{Internet of Things}
\newacronym{ip}{IP}{Internet Protocol}
\newacronym{itu}{ITU}{International Telecommunication Union}
\newacronym{kpi}{KPI}{Key Performance Indicator}
\newacronym{kpm}{KPM}{Key Performance Measurement}
\newacronym{kvm}{KVM}{Kernel-based Virtual Machine}
\newacronym{los}{LOS}{Line-of-Sight}
\newacronym{lsm}{LSM}{Link-to-System Mapping}
\newacronym{lstm}{LSTM}{Long Short Term Memory}
\newacronym{lte}{LTE}{Long Term Evolution}
\newacronym{lxc}{LXC}{Linux Container}
\newacronym{m2m}{M2M}{Machine to Machine}
\newacronym{mac}{MAC}{Medium Access Control}
\newacronym{manet}{MANET}{Mobile Ad Hoc Network}
\newacronym{mano}{MANO}{Management and Orchestration}
\newacronym{mc}{MC}{Multi-Connectivity}
\newacronym{mcc}{MCC}{Mobile Cloud Computing}
\newacronym{mchem}{MCHEM}{Massive Channel Emulator}
\newacronym{mcs}{MCS}{Modulation and Coding Scheme}
\newacronym{mec}{MEC}{Multi-access Edge Computing}
\newacronym{mec2}{MEC}{Mobile Edge Cloud}
\newacronym{mfc}{MFC}{Mobile Fog Computing}
\newacronym{mgen}{MGEN}{Multi-Generator}
\newacronym{mi}{MI}{Mutual Information}
\newacronym{mib}{MIB}{Master Information Block}
\newacronym{miesm}{MIESM}{Mutual Information Based Effective SINR}
\newacronym{mimo}{MIMO}{Multiple Input, Multiple Output}
\newacronym{ml}{ML}{Machine Learning}
\newacronym{mlr}{MLR}{Maximum-local-rate}
\newacronym[plural=\gls{mme}s,firstplural=Mobility Management Entities (MMEs)]{mme}{MME}{Mobility Management Entity}
\newacronym{mmtc}{mMTC}{Massive Machine-Type Communications}
\newacronym{mmwave}{mmWave}{millimeter wave}
\newacronym{mpdccp}{MP-DCCP}{Multipath Datagram Congestion Control Protocol}
\newacronym{mptcp}{MPTCP}{Multipath TCP}
\newacronym{mr}{MR}{Maximum Rate}
\newacronym{mrdc}{MR-DC}{Multi \gls{rat} \gls{dc}}
\newacronym{mse}{MSE}{Mean Square Error}
\newacronym{mss}{MSS}{Maximum Segment Size}
\newacronym{mt}{MT}{Mobile Termination}
\newacronym{mtd}{MTD}{Machine-Type Device}
\newacronym{mtu}{MTU}{Maximum Transmission Unit}
\newacronym{mumimo}{MU-MIMO}{Multi-user \gls{mimo}}
\newacronym{mvno}{MVNO}{Mobile Virtual Network Operator}
\newacronym{nalu}{NALU}{Network Abstraction Layer Unit}
\newacronym{nas}{NAS}{Non-Access Stratum}
\newacronym{nbiot}{NB-IoT}{Narrow Band IoT}
\newacronym{neo}{NEO}{Network Operations}
\newacronym{nfv}{NFV}{Network Function Virtualization}
\newacronym{nfvi}{NFVI}{Network Function Virtualization Infrastructure}
\newacronym{ngrg}{nGRG}{next Generation Research Group}
\newacronym{ni}{NI}{Network Interfaces}
\newacronym{nic}{NIC}{Network Interface Card}
\newacronym{nlos}{NLOS}{Non-Line-of-Sight}
\newacronym{now}{NOW}{Non Overlapping Window}
\newacronym{nsm}{NSM}{Network Service Mesh}
\newacronym{nr}{NR}{New Radio}
\newacronym{nrf}{NRF}{Network Repository Function}
\newacronym{nef}{NEF}{Network Exposure Function}
\newacronym{nsa}{NSA}{Non Stand Alone}
\newacronym{nse}{NSE}{Network Slicing Engine}
\newacronym{nssf}{NSSF}{Network Slice Selection Function}
\newacronym{o2i}{O2I}{Outdoor to Indoor}
\newacronym{oai}{OAI}{OpenAirInterface}
\newacronym{oaicn}{OAI-CN}{\gls{oai} \acrlong{cn}}
\newacronym{oairan}{OAI-RAN}{\acrlong{oai} \acrlong{ran}}
\newacronym{oam}{OAM}{Operations, Administration and Maintenance}
\newacronym{ofdm}{OFDM}{Orthogonal Frequency Division Multiplexing}
\newacronym{olia}{OLIA}{Opportunistic Linked Increase Algorithm}
\newacronym{omec}{OMEC}{Open Mobile Evolved Core}
\newacronym{onap}{ONAP}{Open Network Automation Platform}
\newacronym{onf}{ONF}{Open Networking Foundation}
\newacronym{onos}{ONOS}{Open Networking Operating System}
\newacronym{oom}{OOM}{\gls{onap} Operations Manager}
\newacronym{opnfv}{OPNFV}{Open Platform for \gls{nfv}}
\newacronym{oran}{O-RAN}{Open Radio Access Network}
\newacronym{orbit}{ORBIT}{Open-Access Research Testbed for Next-Generation Wireless Networks}
\newacronym{os}{OS}{Operating System}
\newacronym{oss}{OSS}{Operations Support System}
\newacronym{otic}{OTIC}{Open Testing \& Integration Centre}
\newacronym{pa}{PA}{Position-aware}
\newacronym{pase}{PASE}{Prioritization, Arbitration, and Self-adjusting Endpoints}
\newacronym{pawr}{PAWR}{Platforms for Advanced Wireless Research}
\newacronym{pbch}{PBCH}{Physical Broadcast Channel}
\newacronym{pcef}{PCEF}{Policy and Charging Enforcement Function}
\newacronym{pcfich}{PCFICH}{Physical Control Format Indicator Channel}
\newacronym{pcrf}{PCRF}{Policy and Charging Rules Function}
\newacronym{pdcch}{PDCCH}{Physical Downlink Control Channel}
\newacronym{pdcp}{PDCP}{Packet Data Convergence Protocol}
\newacronym{pdsch}{PDSCH}{Physical Downlink Shared Channel}
\newacronym{pdu}{PDU}{Packet Data Unit}
\newacronym{pf}{PF}{Proportional Fair}
\newacronym{pgw}{PGW}{Packet Gateway}
\newacronym{phich}{PHICH}{Physical Hybrid ARQ Indicator Channel}
\newacronym{phy}{PHY}{Physical}
\newacronym{pmch}{PMCH}{Physical Multicast Channel}
\newacronym{pmi}{PMI}{Precoding Matrix Indicators}
\newacronym{powder}{POWDER}{Platform for Open Wireless Data-driven Experimental Research}
\newacronym{ppo}{PPO}{Proximal Policy Optimization}
\newacronym{ppp}{PPP}{Poisson Point Process}
\newacronym{prach}{PRACH}{Physical Random Access Channel}
\newacronym{prb}{PRB}{Physical Resource Block}
\newacronym{psnr}{PSNR}{Peak Signal to Noise Ratio}
\newacronym{pss}{PSS}{Primary Synchronization Signal}
\newacronym{pucch}{PUCCH}{Physical Uplink Control Channel}
\newacronym{pusch}{PUSCH}{Physical Uplink Shared Channel}
\newacronym{rar}{RAR}{Random Access Response}
\newacronym{qam}{QAM}{Quadrature Amplitude Modulation}
\newacronym{qci}{QCI}{\gls{qos} Class Identifier}
\newacronym{5qi}{5QI}{5G \gls{qos} Identifier}
\newacronym{qoe}{QoE}{Quality of Experience}
\newacronym{qos}{QoS}{Quality of Service}
\newacronym{quic}{QUIC}{Quick UDP Internet Connections}
\newacronym{sfp}{SFP}{Small Form-factor Pluggable}
\newacronym{rach}{RACH}{Random Access Channel}
\newacronym{ran}{RAN}{Radio Access Network}
\newacronym[firstplural=Radio Access Technologies (RATs)]{rat}{RAT}{Radio Access Technology}
\newacronym{rcn}{RCN}{Research Coordination Network}
\newacronym{rc}{RC}{RAN Control}
\newacronym{rec}{REC}{Radio Edge Cloud}
\newacronym{red}{RED}{Random Early Detection}
\newacronym{renew}{RENEW}{Reconfigurable Eco-system for Next-generation End-to-end Wireless}
\newacronym{rf}{RF}{Radio Frequency}
\newacronym{rfc}{RFC}{Request for Comments}
\newacronym{rfr}{RFR}{Random Forest Regressor}
\newacronym{ric}{RIC}{RAN Intelligent Controller}
\newacronym{rlc}{RLC}{Radio Link Control}
\newacronym{rlf}{RLF}{Radio Link Failure}
\newacronym{rlnc}{RLNC}{Random Linear Network Coding}
\newacronym{rmr}{RMR}{RIC Message Router}
\newacronym{rmse}{RMSE}{Root Mean Squared Error}
\newacronym{rnis}{RNIS}{Radio Network Information Service}
\newacronym{rr}{RR}{Round Robin}
\newacronym{rrc}{RRC}{Radio Resource Control}
\newacronym{rrm}{RRM}{Radio Resource Management}
\newacronym{rru}{RRU}{Remote Radio Unit}
\newacronym{rs}{RS}{Remote Server}
\newacronym{rsrp}{RSRP}{Reference Signal Received Power}
\newacronym{rsrq}{RSRQ}{Reference Signal Received Quality}
\newacronym{rss}{RSS}{Received Signal Strength}
\newacronym{rssi}{RSSI}{Received Signal Strength Indicator}
\newacronym{rtt}{RTT}{Round Trip Time}
\newacronym{ru}{RU}{Radio Unit}
\newacronym{rw}{RW}{Receive Window}
\newacronym{rx}{RX}{Receiver}
\newacronym{s1ap}{S1AP}{S1 Application Protocol}
\newacronym{sa}{SA}{standalone}
\newacronym{sack}{SACK}{Selective Acknowledgment}
\newacronym{sap}{SAP}{Service Access Point}
\newacronym{sc2}{SC2}{Spectrum Collaboration Challenge}
\newacronym{scef}{SCEF}{Service Capability Exposure Function}
\newacronym{sch}{SCH}{Secondary Cell Handover}
\newacronym{scoot}{SCOOT}{Split Cycle Offset Optimization Technique}
\newacronym{sctp}{SCTP}{Stream Control Transmission Protocol}
\newacronym{sdap}{SDAP}{Service Data Adaptation Protocol}
\newacronym{sdk}{SDK}{Software Development Kit}
\newacronym{sdm}{SDM}{Space Division Multiplexing}
\newacronym{sdma}{SDMA}{Spatial Division Multiple Access}
\newacronym{sdn}{SDN}{Software-defined Networking}
\newacronym{sdr}{SDR}{Software-defined Radio}
\newacronym{seba}{SEBA}{SDN-Enabled Broadband Access}
\newacronym{sgsn}{SGSN}{Serving GPRS Support Node}
\newacronym{sgw}{SGW}{Service Gateway}
\newacronym{si}{SI}{Study Item}
\newacronym{sib}{SIB}{Secondary Information Block}
\newacronym{sinr}{SINR}{Signal to Interference plus Noise Ratio}
\newacronym{sip}{SIP}{Session Initiation Protocol}
\newacronym{siso}{SISO}{Single Input, Single Output}
\newacronym{sla}{SLA}{Service Level Agreement}
\newacronym{sm}{SM}{Service Model}
\newacronym{smf}{SMF}{Session Management Function}
\newacronym{smo}{SMO}{Service Management and Orchestration}
\newacronym{sms}{SMS}{Short Message Service}
\newacronym{smsgmsc}{SMS-GMSC}{\gls{sms}-Gateway}
\newacronym{snr}{SNR}{Signal-to-Noise-Ratio}
\newacronym{cnr}{CNR}{Carrier-to-Noise-Ratio}
\newacronym{son}{SON}{Self-Organizing Network}
\newacronym{sptcp}{SPTCP}{Single Path TCP}
\newacronym{srb}{SRB}{Service Radio Bearer}
\newacronym{srn}{SRN}{Standard Radio Node}
\newacronym{srs}{SRS}{Sounding Reference Signal}
\newacronym{prs}{PRS}{Positioning Reference Signal}
\newacronym{zc}{ZC}{Zadoff-Chu}
\newacronym{ta}{TA}{Timing Advance}
\newacronym{ss}{SS}{Synchronization Signal}
\newacronym{ssb}{SSB}{Synchronization Signal Block}
\newacronym{sss}{SSS}{Secondary Synchronization Signal}
\newacronym{st}{ST}{Spanning Tree}
\newacronym{svc}{SVC}{Scalable Video Coding}
\newacronym{tb}{TB}{Transport Block}
\newacronym{tcp}{TCP}{Transmission Control Protocol}
\newacronym{tdd}{TDD}{Time Division Duplexing}
\newacronym{tdm}{TDM}{Time Division Multiplexing}
\newacronym{tdma}{TDMA}{Time Division Multiple Access}
\newacronym{tfl}{TfL}{Transport for London}
\newacronym{tfrc}{TFRC}{TCP-Friendly Rate Control}
\newacronym{tft}{TFT}{Traffic Flow Template}
\newacronym{tgen}{TGEN}{Traffic Generator}
\newacronym{tip}{TIP}{Telecom Infra Project}
\newacronym{tm}{TM}{Transparent Mode}
\newacronym{to}{TO}{Telco Operator}
\newacronym{tr}{TR}{Technical Report}
\newacronym{trp}{TRP}{Transmitter Receiver Pair}
\newacronym{ts}{TS}{Technical Specification}
\newacronym{tti}{TTI}{Transmission Time Interval}
\newacronym{ttt}{TTT}{Time-to-Trigger}
\newacronym{tx}{TX}{Transmitter}
\newacronym{uas}{UAS}{Unmanned Aerial System}
\newacronym{uav}{UAV}{Unmanned Aerial Vehicle}
\newacronym{udm}{UDM}{Unified Data Management}
\newacronym{udp}{UDP}{User Datagram Protocol}
\newacronym{udr}{UDR}{Unified Data Repository}
\newacronym{ue}{UE}{User Equipment}
\newacronym{uhd}{UHD}{\gls{usrp} Hardware Driver}
\newacronym{ul}{UL}{Uplink}
\newacronym{um}{UM}{Unacknowledged Mode}
\newacronym{uml}{UML}{Unified Modeling Language}
\newacronym{upa}{UPA}{Uniform Planar Array}
\newacronym{ula}{ULA}{Uniform Linear Array}
\newacronym{upf}{UPF}{User Plane Function}
\newacronym{urllc}{URLLC}{Ultra Reliable and Low Latency Communications}
\newacronym{usa}{U.S.}{United States}
\newacronym{usim}{USIM}{Universal Subscriber Identity Module}
\newacronym{usrp}{USRP}{Universal Software Radio Peripheral}
\newacronym{utc}{UTC}{Urban Traffic Control}
\newacronym{vim}{VIM}{Virtualization Infrastructure Manager}
\newacronym{vm}{VM}{Virtual Machine}
\newacronym{vnf}{VNF}{Virtual Network Function}
\newacronym{volte}{VoLTE}{Voice over \gls{lte}}
\newacronym{voltha}{VOLTHA}{Virtual OLT HArdware Abstraction}
\newacronym{vr}{VR}{Virtual Reality}
\newacronym{vran}{vRAN}{Virtualized \gls{ran}}
\newacronym{vss}{VSS}{Video Streaming Server}
\newacronym{v2x}{V2X}{Vehicle-to-Everything}
\newacronym{wbf}{WBF}{Wired Bias Function}
\newacronym{wf}{WF}{Waterfilling}
\newacronym{wg}{WG}{Working Group}
\newacronym{wlan}{WLAN}{Wireless Local Area Network}
\newacronym{osm}{OSM}{Open Source Management and Orchestration}
\newacronym{pnf}{PNF}{Physical Network Function}
\newacronym{drl}{DRL}{Deep Reinforcement Learning}
\newacronym{mtc}{MTC}{Machine-type Communications}
\newacronym{osc}{OSC}{O-RAN Software Community}
\newacronym{mns}{MnS}{Management Services}
\newacronym{ves}{VES}{\gls{vnf} Event Stream}
\newacronym{ei}{EI}{Enrichment Information}
\newacronym{fh}{FH}{Fronthaul}
\newacronym{fft}{FFT}{Fast Fourier Transform}
\newacronym{laa}{LAA}{Licensed-Assisted Access}
\newacronym{plfs}{PLFS}{Physical Layer Frequency Signals}
\newacronym{ptp}{PTP}{Precision Time Protocol}
\newacronym{asic}{ASIC}{Application-specific Integrated Circuit}
\newacronym{aal}{AAL}{Acceleration Abstraction Layer}
\newacronym{fec}{FEC}{Forward Error Correction}
\newacronym{sdl}{SDL}{Shared Data Layer}
\newacronym{nib}{NIB}{Network Information Base}
\newacronym{rnib}{R-NIB}{RAN \gls{nib}}
\newacronym{fcaps}{FCAPS}{Fault, Configuration, Accounting, Performance, Security}
\newacronym{ie}{IE}{Information Element}
\newacronym{fg}{FG}{Focus Group}
\newacronym{osfg}{OSFG}{Open Source Focus Group}
\newacronym{sdfg}{SDFG}{Standard Development Focus Group}
\newacronym{tifg}{TIFG}{Test \& Integration Focus Group}
\newacronym{sfg}{SFG}{Security Focus Group}
\newacronym{swg}{SWG}{Security Work Group}
\newacronym{e2sm}{E2SM}{E2 Service Model}
\newacronym{tsc}{TSC}{Technical Steering Committee}
\newacronym{sdo}{SDO}{Standard-Development Organization}
\newacronym{sql}{SQL}{Structured Query Language}
\newacronym{ssh}{SSH}{Secure Shell}
\newacronym{tls}{TLS}{Transport Layer Security}
\newacronym{netconf}{NETCONF}{Network Configuration Protocol}
\newacronym{dtls}{DTLS}{Datagram Transport Layer Security}
\newacronym{cmp}{CMP}{Certificate Management Protocol}
\newacronym{ccc}{CCC}{Cell Configuration and Control}
\newacronym{dsp}{DSP}{Digital Signal Processing}
\newacronym{opex}{OPEX}{Operational Expenses}
\newacronym{cbrs}{CBRS}{Citizen Broadband Radio Service}
\newacronym{ntn}{NTN}{Non-terrestrial Network}
\newacronym{gbr}{GBR}{Guaranteed Bitrate}
\newacronym{sps}{SPS}{Semi-Persistent Scheduling}
\newacronym{tbs}{TBS}{Transport Block Size}
\newacronym{gnss}{GNSS}{Global Navigation Satellite System}
\newacronym{tof}{ToF}{Time of Flight}
\newacronym{rtof}{RToF}{Return Time of Flight}
\newacronym{rsig}{RS}{Reference Signal}
\newacronym{nrtric}{near-RT RIC}{near-Real Time RAN Intelligent Controller}
\newacronym{nonrtric}{non-RT RIC}{non-Real Time RAN Intelligent Controller}
\newacronym{ngran}{NG-RAN}{Next-Generation Radio Access Network}
\newacronym{rics}{RICs}{RAN Intelligent Controllers}
\newacronym{aoa}{AoA}{Angle of Arrival}
\newacronym{aod}{AoD}{Angle of Departure}
\newacronym{tdoa}{TDoA}{Time Difference of Arrival}
\newacronym{rtoa}{RToA}{Return Time of Arrival}
\newacronym{ecdf}{ECDF}{Empirical Cumulative Distribution Function}
\newacronym{ris}{RIS}{Reconfigurable Intelligent Surface}
\newacronym{srd}{SRD}{Smart Radio Device}
\newacronym{lcs}{LCS}{Location Services}
\newacronym{gfbr}{GFBR}{Guaranteed Flow Bit Rate}
\newacronym{rg}{RG}{Resource Grid}
\newacronym{rb}{RB}{Resource Block}
\newacronym{rbs}{RBs}{Resource Blocks}
\newacronym{re}{RE}{Resource Element}
\newacronym{rfra}{RF}{Radio Frame}
\newacronym{scs}{SCS}{Subcarrier Spacing}
\newacronym{ec}{EC}{Edge Computing}
\newacronym{5g-phy}{5G PHY}{5G Physical Layer}
\newacronym{fr1}{FR1}{Frequency Range 1}
\newacronym{tpm}{tpm}{times per minute}
\newacronym{ra}{RA}{Random Access}
\newacronym{crnti}{C-RNTI}{Cell Radio Network Temporary Identifier}
\newacronym{toa}{ToA}{Time of Arrival}
\newacronym{kf}{KF}{Kalman Filter}
\newacronym{lmf}{LMF}{Location Management Function}
\newacronym{nsps}{$N_{symb}^{slot}$}{Number of symbols per slot}
\newacronym{nsysrs}{$N_{symb}^{SRS}$}{Consecutive symbols per SRS subcarrier}
\newacronym{nspf}{$N_{slot}^{frame,\mu}$}{Number of slots per frame}
\newacronym{nspsf}{$N_{slot}^{subframe,\mu}$}{Number of slots per subframe}   %
\newacronym{nscprb}{$N_{sc}^{RB}$}{Number of subcarriers per resource block}
\newacronym{msrs}{$m_{SRS,b}$}{Number of PRBs per SRS}
\newacronym{Msrs}{$M_{sc,b}^{SRS}$}{Number of subcarriers per SRS}
\newacronym{ktc}{$k_{TC}$}{Trasmission Comb Constant}
\newacronym{inioff}{$l_{offset}$}{Initial Offset}
\newacronym{dn}{DN}{Data Network}
\newacronym{sba}{SBA}{Service-Based Architecture}
\newacronym{seaf}{SEAF}{Security Anchor Functionality}
\newacronym{pcf}{PCF}{Policy Control Function}
\newacronym{nf}{NF}{Network Function}
\newacronym{crlb}{CRLB}{Cramér–Rao Lower Bound}
\newacronym{music}{MUSIC}{MUltiple SIgnal Classification}
\newacronym{esprit}{ESPRIT}{Estimation of Signal Parameters via Rotational Invariance Techniques}
\newacronym{iaa}{IAA}{Iterative Adaptive Approaches}
\newacronym{samv}{SAMV}{Sparse Asymptotic Minimum Variance}
\newacronym{sml}{SML}{Stochastic Machine Learning}
\newacronym{em}{EM}{Expectation-Maximization}
\newacronym{lo}{LO}{Local Oscillator}
\newacronym{ota}{OTA}{over-the-air}
\newacronym{jade}{JADE}{Joint Angle and Delay Estimation}

\def\BibTeX{{\rm B\kern-.05em{\sc i\kern-.025em b}\kern-.08em
    T\kern-.1667em\lower.7ex\hbox{E}\kern-.125emX}}
\begin{document}

\title{AoA Services in 5G Networks:\\A Framework for Real-World Implementation and Systematic Testing
}

\author{\IEEEauthorblockN{Alberto Ceresoli, Viola Bernazzoli, Roberto Pegurri, Ilario Filippini}
\IEEEauthorblockA{
\textit{Politecnico di Milano}, Milan, Italy \\
\textit{name}.\textit{surname}@polimi.it}\vspace{-1cm}
}

\maketitle

\begin{abstract}
%
Accurate positioning is a key enabler for emerging 5G applications. While the standardized Location Management Function (LMF) operates centrally within the core network, its scalability and latency limitations hinder low-latency and fine-grained localization. A practical alternative is to shift positioning intelligence toward the radio access network (RAN), where uplink sounding reference signal (SRS)-based angle-of-arrival (AoA) estimation offers a lightweight, network-native solution. In this work, we present the first fully open-source 5G testbed for AoA estimation, enabling systematic and repeatable experimentation under realistic yet controllable channel conditions. The framework integrates the NVIDIA Sionna RT with a Keysight PROPSIM channel emulator and includes a novel phase calibration procedure for USRP~N310 devices. Experimental results show sub-degree to few-degree accuracy, validating the feasibility of lightweight, single-anchor, network-native localization within next-generation 5G systems.
\end{abstract}

\begin{IEEEkeywords}
Channel emulator, Sionna RT, Angle-of-Arrival, 5G \vspace{-0.5cm}
\end{IEEEkeywords}

\section{Introduction}%
\label{sec: intro}




The rapid evolution of wireless communication systems has brought about new demands not only in terms of data throughput and latency but also in high-precision positioning, especially for use cases such as autonomous vehicles, drone navigation, and \gls{iot} networks \cite{iot_survey}. Traditional positioning techniques, such as \gls{gnss}, often suffer from limited coverage or poor accuracy in indoor, urban canyon, or densely populated environments \cite{gnss_perform}.

To bridge this gap, 3GPP standardization bodies have foreseen, for the first time in \gls{5g} releases, the positioning service as a fundamental feature of the cellular network. Specifically, they introduced the \gls{lmf} as one of the \glspl{vnf} that can be instantiated within the \gls{cn}. However, the centralized architecture of the \gls{lmf} introduces scalability and latency constraints, as the \gls{cn} is typically deployed in a limited number of national-level sites. This centralization limits its applicability in scenarios demanding low-latency or high-precision positioning, particularly when rapid adaptation to dynamic radio conditions is required.

A more promising vision is to move positioning intelligence closer to the network edge, embedding it directly within the \gls{ran}. In this context, the architectural innovations introduced by the \gls{oran} Alliance provide a natural enabler \cite{oran_polese}. By exposing open interfaces and programmable control loops, said \glspl{ric}, \gls{oran} allows the development of localization micro-services (xApps) that can seamlessly integrate with the radio stack.

Within this framework, uplink \gls{aoa} estimation based on the \gls{ue}'s \gls{srs} emerges as a particularly attractive primitive. This approach operates transparently to the \gls{ue}, requiring no additional hardware or signaling overhead, and can seamlessly coexist with other network-native features. By confining the computation to the network edge, it enables a fast, lightweight, and easily deployable positioning service. Moreover, when combined with uplink \gls{ue} ranging techniques \cite{bernazzoli2025robust}, this solution supports single-anchor localization, eliminating the need for inter-\gls{gnb} synchronization and orchestration, procedures that are typically complex and costly. Consequently, \gls{aoa} serves as a lightweight yet powerful building block for network-native, single-anchor positioning.

Indeed, several studies (e.g., \cite{joint_li, sun_comparative_single_gnb}) have investigated this promising research direction, primarily through simulation-based evaluations. In contrast, only a few works have addressed the physical deployment and testing of \gls{aoa}-based positioning, as pursued in this paper. For example, the testbeds presented in \cite{5g_aoa_techniques_comparizon} and \cite{xhafa_comparizon} demonstrate the feasibility of \gls{aoa} estimation at the hardware level but are limited to minimal setups operating solely at the physical layer. We introduce the first fully open-source experimental framework that enables unrestricted and systematic testing of a complete \gls{5g} stack, namely, a fully operational 5G network with an integrated \gls{aoa} service supporting both prototype and \gls{cots} \gls{ue} devices. This setup allows us to account for fundamental real-world non-idealities arising from hardware impairments and propagation environment characteristics. Consequently, the proposed testbed provides a flexible and reliable platform for reproducible and extensive performance evaluation of network-native positioning techniques.

In particular, by integrating a scenario-based ray tracer with a commercial channel emulator, we enable the evaluation of 5G positioning services across a wide range of scenarios under realistic and fully repeatable propagation conditions. In parallel, the specific challenge of \gls{aoa} estimation is investigated through an extensive experimental campaign.

\noindent\textbf{Main Contributions.} This work provides the following key contributions. \textit{i)} We present the first fully open-source experimental setup for evaluating positioning techniques in real 5G networks, supporting both prototype and \gls{cots} \glspl{ue}. \textit{ii)} We design and integrate in a full-stack 5G network an uplink \gls{aoa}-based positioning function operating at the network edge and leveraging \gls{ue} \glspl{srs}. \textit{iii)} We combine a scenario-based ray tracer (Sionna RT~\cite{sionna}) with a commercial channel emulator (Keysight PROPSIM) to reproduce realistic and fully repeatable propagation conditions. \textit{iv)} The proposed framework inherently captures hardware and environmental non-idealities, bridging the gap between simulation and practical deployment. \textit{v)} An extensive experimental campaign validates the feasibility and performance of uplink \gls{aoa}-based positioning in real 5G conditions.

\noindent\textbf{Paper structure.} We organize the rest of the paper as follows: 
Section~\ref{sec: back} provides the fundamental background, Section~\ref{sec: method} dives into the details of the \gls{aoa} estimation method, Section~\ref{sec: setup} describes the testbed we built for validating the method, Section~\ref{sec: results} details our extensive experimental campaign, and last, Section~\ref{sec: concl} summarizes the findings.

\section{Background}%
\label{sec: back}

\subsection{5G physical layer}
The \gls{5g} \gls{nr} physical layer is built on a scalable \gls{ofdm}-based resource grid that adapts to diverse service requirements. Time is organized into radio frames of 10 ms, each divided into subframes, slots, \glspl{rb}, and OFDM symbols. The \gls{nspsf} depends on the numerology index $\mu$, which further influences the subcarrier spacing, defined as $15 \cdot 2^{\mu}$ kHz. Within this grid, user data and control signaling are jointly mapped to subframes, slots, and subcarriers, enabling flexible allocation of radio resources with adjustable trade-offs between signaling overhead and accuracy. This flexible numerology allows for balancing throughput, latency, and coverage requirements. 
Moreover, \gls{5g} supports advanced multi-antenna techniques such as \gls{mimo} and beamforming, providing the spatial resolution needed to extract \gls{aoa} information.

\subsection{Sounding Reference Signal (SRS)} 
Among the various reference signals defined in \gls{5g} \gls{nr}, the uplink \gls{srs} plays a central role in this work. Originally designed for channel sounding, uplink scheduling, and beam management, the \gls{srs} is a deterministic, unmodulated sequence transmitted by the \gls{ue} and known at the \gls{gnb} \cite{38211}. Its structure is based on Zadoff–Chu sequences, characterized by constant amplitude and zero autocorrelation properties, ensuring high robustness to interference and make them well suited for both channel estimation and positioning applications.

The \gls{srs} can be flexibly configured in both time and frequency domains. It may occupy one or more contiguous \gls{ofdm} symbols within a slot and can be mapped across interleaved \glspl{rb} over a configurable bandwidth. This flexibility enables the multiplexing of multiple \glspl{ue} and promotes efficient reuse of radio resources. The network’s knowledge of the \gls{srs} sequence and its allocation in the resource grid makes it particularly suitable for uplink-based localization, allowing joint \gls{toa} and \gls{aoa} estimation without any hardware or software modifications to the \gls{ue}. In this work, we exploit \gls{srs} transmissions to perform \gls{aoa} estimation using subspace-based algorithms.



\subsection{Angle-of-Arrival (AoA) techniques}
The potential of \gls{aoa}-based techniques was first explored in the 1990s by \cite{jade} through the development of the \gls{jade} family of algorithms, which enable the joint extraction of \gls{aoa} and \gls{tof}. However, the integration of such techniques into mobile \glspl{ran} has attracted significant attention only in the past decade. Notable contributions have focused on \gls{lte} systems \cite{joint_LTE}, while the architectural innovations introduced with \gls{5g} \gls{ran} have further revitalized interest in this field. Despite their theoretical effectiveness, \gls{jade}-based algorithms are computationally demanding, mainly due to the eigendecomposition of large covariance matrices and the associated two-dimensional spectral search. Consequently, recent research efforts have shifted toward the adoption of more lightweight and robust \gls{aoa} and \gls{toa} estimation methods leveraging native \gls{5g} signaling.

\section{AoA Estimation Framework}%
\label{sec: method}


In this section, we present the framework used to estimate the \glspl{aoa} of connected \glspl{ue} from a serving \gls{gnb}. In the proposed setup, a single anchor --- namely, the \gls{gnb} --- is employed, equipped with an $M$-element uniform linear array (\gls{ula}). However, the proposed method can be readily extended and integrated with \gls{tof}-based ranging techniques \cite{bernazzoli2025robust} to enable single-anchor localization.


\subsection{AoA estimation}\label{subsec: aoa}
To further assess the systematic capability of the testbed, we decided to test the performance of two different subspace-based techniques under the same environment and channel conditions. 
These techniques are based on the observation that the array covariance matrix can be decomposed into orthogonal signal and noise subspaces. The steering vectors corresponding to true \gls{aoa} lie in the signal subspace, hence resulting in an orthogonal direction to the noise subspace.

To model the received signal, we assume a \gls{ula} of $M$ antenna elements, inter-spaced by $\Delta d=\lambda/2$, ensuring coverage over the range $[-90^\circ, +90^\circ]$. Under the narrowband and far-field assumptions, which are reasonable for \gls{srs} and \gls{5g} \gls{ran} deployments, a source impinging from angle $\theta$ introduces a deterministic phase shift between adjacent antennas, captured by the steering vector
\begin{equation}
\label{eq:steer_vect}
\mathbf{a}(\theta) =
\begin{bmatrix}
1 & e^{j\mu} & e^{j2\mu} & \cdots & e^{j(M-1)\mu}
\end{bmatrix}^T
\end{equation}
where $\mu = -\dfrac{2\pi}{\lambda}\Delta d\cdot\sin(\theta)$ is the spatial frequency, and $T$ represents the transpose operator.

The \gls{ula} received signals at time $t$ are thus expressed as
\begin{equation}
\label{eq:signal_model}
\mathbf{x}(t) = \mathbf{a}(\theta) s(t) + \mathbf{n}(t),
\end{equation}
where $s(t)$ denotes the transmitted signal after propagation attenuation, and $\mathbf{n}(t)$ represents an additive white Gaussian noise (AWGN) vector. In our case, $s(t)$ corresponds to the uplink \gls{srs}. Since \gls{srs} transmissions are orthogonal across users, the single-source model is valid, thereby simplifying the estimation process. For clarity, the following mathematical formulation focuses on the single-source case; however, the same derivations can be extended to multi-source \gls{aoa} estimation when the number of sources is known.

The covariance matrix $\mathbf{R}_x$ captures the spatial correlation among the signals received by the $M$ antenna elements and is defined as
\begin{equation}
\label{eq:Rx_paper}
\mathbf{R}_x = \mathbb{E}\{\mathbf{x}(t)\mathbf{x}^H(t)\}
= \mathbf{a}(\theta)\mathbf{R}_s\mathbf{a}^H(\theta) + \sigma^2 \mathbf{I},
\end{equation}
where $\mathbf{x}(t)$ is the received signal vector across the antenna array at time~$t$, and $\mathbf{x}^H(t)$ denotes its Hermitian transpose. The operator $\mathbb{E}{\cdot}$ represents the statistical expectation, while $\mathbf{R}_s \triangleq \mathbb{E}{\mathbf{s}(t)\mathbf{s}^H(t)}$ is the source signal covariance matrix. The noise is assumed to be zero-mean \gls{awgn} with covariance $\sigma^2 = \mathbb{E}{\mathbf{n}(t)\mathbf{n}^H(t)}$, and $\mathbf{I}$ denotes the identity matrix.

In practice, the covariance matrix is estimated from the phase-corrected received samples as
\begin{equation}
\label{eq:Rx_discrete_paper}
\hat{\mathbf{R}}_x = \frac{1}{M_{sc,b}^{\text{SRS}}}\sum_{n=0}^{M_{sc,b}^{\text{SRS}}-1} \mathbf{x'}[n]\mathbf{x'}^H[n],
\end{equation}
where $M_{sc,b}^{\text{SRS}}$ denotes the number of \gls{srs} samples within the considered bandwidth, and $\mathbf{x}'[n]$ represents the $n$-th phase-corrected \gls{srs} sample received at the \gls{ula}. By performing the eigendecomposition of  $\hat{\mathbf{R}}_{x} \triangleq \mathbf{V} \boldsymbol{\Lambda} \mathbf{V}^H$, we get $\boldsymbol{\Lambda}=diag\{\lambda_0,\dots,\lambda_M\}$ containing the sorted eigenvalues, and $\mathbf{V}$ their associated eigenvectors, of which the leftmost $M-1$ correspond to the noise subspace, while the remaining ones define the signal subspace $\mathbf{V}_s$.

\subsubsection{MUSIC}
The \acrlong{music} algorithm \cite{music} offers higher resolution compared to classical beamforming techniques \cite{aoa_bible}, as it enables the discrimination of closely spaced multipath components even under limited \gls{snr} conditions.
 
The algorithm exploits the orthogonality condition between the signal and noise subspaces, expressed as
\begin{equation}\label{eq: orthogonality}
\mathbf{a}^H(\theta) \mathbf{V}_n \mathbf{V}_n^H \mathbf{a}(\theta) = 0,
\end{equation}
where $\mathbf{V}n$ denotes the matrix of noise eigenvectors, i.e. the noise subspace of $\hat{\mathbf{R}}_x$. This condition leads to the MUSIC pseudospectrum $P(\theta)$, whose peaks correspond to the estimated angles of arrival $\hat{\theta}$:
\begin{equation}\label{eq: estimated_angle}
\hat{\theta}  = \arg \max_\theta\{P(\theta)\} \\
              = \arg \max_\theta \Bigl\{ \frac{1}{\mathbf{a}^H(\theta)\mathbf{V}_n\mathbf{V}_n^H\mathbf{a}(\theta)} \Bigr\}.
\end{equation}
%

\subsubsection{ESPRIT}
The \acrlong{esprit} outperforms MUSIC in terms of computational and storage requirements by exploiting the shift-invariance property of the array \cite{invariance_shift}.

ESPRIT assumes that the \gls{ula} can be decomposed into two equally-sized (and possibly overlapping) sub-arrays, where $sub_1$ is a $k\Delta d$-shifted version of $sub_0$. The received signal $\mathbf{x}_i(t)$ at each $sub_i$ is modeled as in \eqref{eq:signal_model}:
%
\begin{equation}
    \begin{split}
    \begin{bmatrix}
        \mathbf{x}_0(t) \\ \mathbf{x}_1(t)
    \end{bmatrix}
    =\begin{bmatrix}
        \mathbf{a}(\theta) \\ \mathbf{a}(\theta) \mathbf{\Phi}
    \end{bmatrix}
    \mathbf{s}(t)
    +\begin{bmatrix}
        \mathbf{n}_0(t) \\ \mathbf{n}_1(t)
    \end{bmatrix}
    = \tilde{\mathbf{a}}(\theta) \mathbf{s}(t) + \mathbf{n}(t).
    \end{split}
\end{equation}
where $\mathbf{a}(\theta)$ is the steering vector as in \eqref{eq:steer_vect}. The signal received by $sub_1$ will experience an extra delay due to the shift $k\Delta d$ described by the diagonal matrix 
$\mathbf{\Phi} = [e^{-j\frac{2\pi}{ \lambda}k \Delta d \sin{\theta}}]$. 
The objective of the ESPRIT technique is to infer the \gls{aoa} through the determination of $\mathbf{\Phi}$. To do so, two steps are required: first to estimate the signal subspace $\mathbf{V}_s$ derived from $\hat{\mathbf{R}}_x$ as seen in \eqref{eq:Rx_discrete_paper}, then to extract 
$\mathbf{\Phi}$.

Once $\mathbf{V}_s$ is obtained, it can be decomposed into $\mathbf{V}_{s0}$ and $\mathbf{V}_{s1}$ for $sub_0$ and $sub_1$ respectively, to form the invariance equation $\mathbf{V}_{s0} \mathbf{\Psi} = \mathbf{V}_{s1}$. This equation lets us estimate the subspace rotating operator $\hat{\mathbf{\Psi}}$ by means of least-squares technique. From $\hat{\mathbf{\Psi}}$, thanks to the shift-invariance property, we can compute $\hat{\mathbf{\Phi}}$ by means of eigendecomposition:
\begin{equation}
\label{eq:psidecom}
\hat{\mathbf{\Psi}} = \mathbf{T} \hat{\mathbf{\Phi}} \mathbf{T}^{-1}, \;\
\exists\, \mathbf{T} \ \text{s.t.} \ ( \det\mathbf{T} \neq 0  \;\wedge\; \mathbf{a}(\theta) = \mathbf{V}_s \mathbf{T}).
\end{equation}
%
Eigenvalues of \eqref{eq:psidecom} allow the \gls{aoa} extraction from $\hat{\mathbf{\Phi}}$ via:
\begin{equation}
    \hat \theta = \arcsin \biggl( - \frac{\lambda}{2 \pi k \Delta d} \mu \biggr).
\end{equation}

In the proposed uplink \gls{srs}-based configuration, MUSIC and ESPRIT approach provide high-resolution user direction estimation with moderate computational complexity, making it suitable for potential real-time deployment.

\subsection{Calibration}\label{subsec: calibration}

\begin{figure}
    \centering
    \includegraphics[trim={0 20 0 20},clip,width=\linewidth]{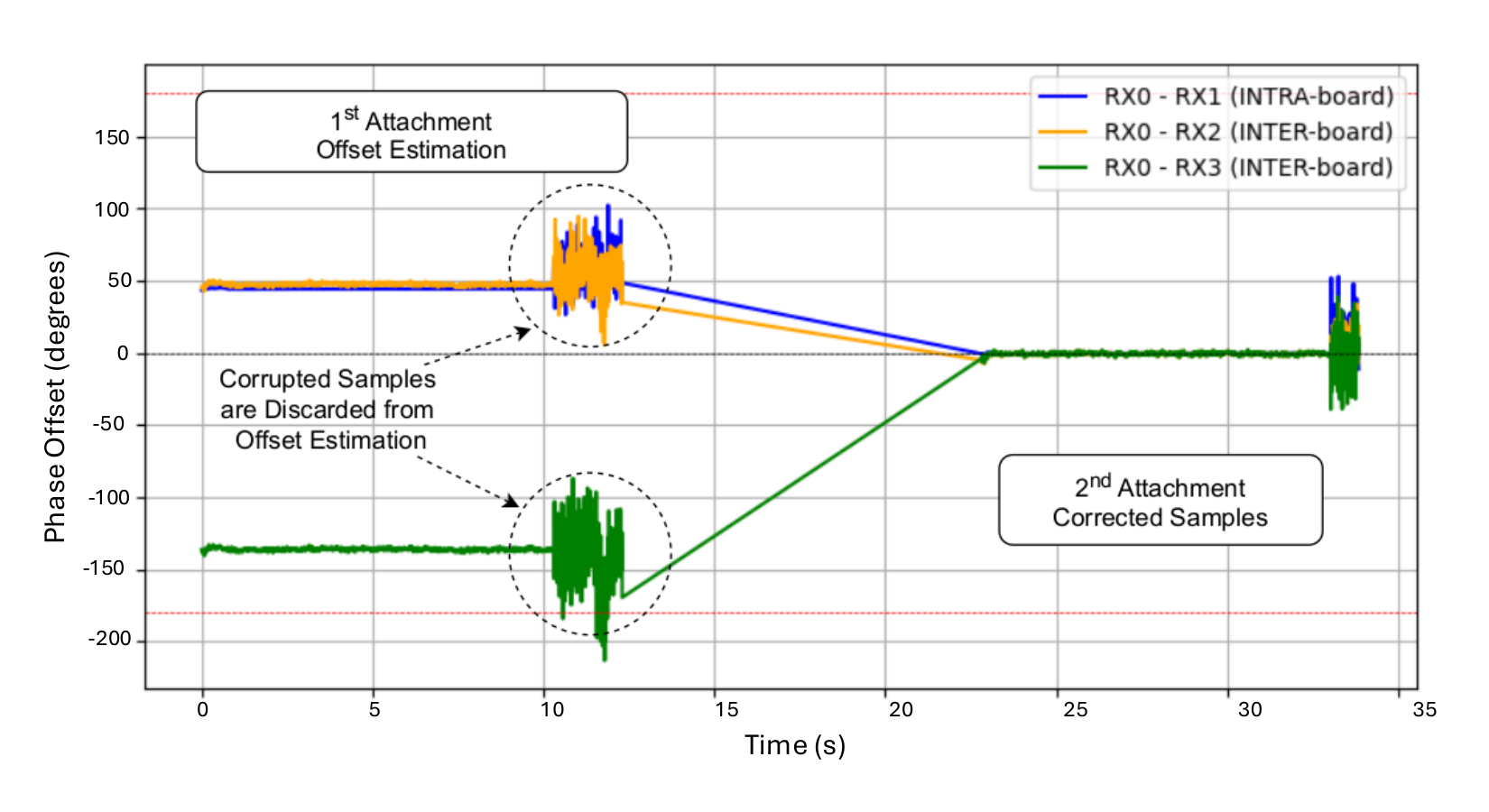} 
    \caption{\small Calibration example, by using a 4-element \gls{ula}.}
    \label{fig: phases}
\end{figure}

One of the main challenges in real-world \gls{aoa} estimation is the hardware-dependent phase misalignment among antenna ports. In our setup, the employed \gls{sdr} is equipped with two separate daughterboards, each featuring an independent \gls{rf} chain and local oscillator. As stated in the manufacturer’s documentation:
\begin{quote}
\textit{“The N310 has no way of aligning the phase between channels, and it will be random between runs.”}~\cite{n310_user}
\end{quote}
As a result, the relative carrier phases are not inherently synchronized at startup, introducing an arbitrary phase offset between antenna ports. Without proper correction, this offset prevents accurate \gls{aoa} estimation by algorithms that rely on inter-antenna phase differences.

To ensure that the inter-antenna phase difference remains stable for the duration of each measurement campaign, we modeled a first \gls{ota} calibration mechanism, performed opportunistically at system start-up.

A reference \gls{ue} is temporarily positioned at $0^\circ$ in front of the array, where the theoretical inter-antenna phase difference is zero under narrowband and far-field assumptions, conditions satisfied by the transmitted \gls{srs} and the \gls{ue} placement. From the received uplink \gls{srs} sequences, the instantaneous phase offsets of each antenna port relative to the reference port 0, denoted $\Delta\hat{\phi}_{0,m}$, where $m \in (0, M]$, are estimated and averaged over a short time interval to obtain $\Delta\bar{\phi}_{0,m}$. These offsets represent per-port phase biases that remain constant until the next \gls{sdr} reboot and are stored for the duration of the measurement campaign.

During test operations, each received sample $x_m(n)$ from port $m$ is thus digitally corrected as
\begin{equation}
x_m'(n) = x_m(n) \cdot e^{-j \Delta\bar{\phi}_{0,m}} \quad for \; n\in[0, M_{sc,b}^{\text{SRS}}-1],
\end{equation}
where $M_{sc,b}^{\text{SRS}}$ denotes the length of the \gls{srs} sequence, corresponding to $960$ samples in the considered full-bandwidth configuration at $60$MHz. This procedure ensures consistent phase alignment across antennas during multiple UE attachment and detachment cycles. 

This method requires no external reference or additional hardware, embedding the calibration into normal network operation and ensuring stable inter-antenna phase coherence for \gls{aoa} estimation.

\subsection{Cylindrical Correction}

\begin{figure}
    \centering 
    \includegraphics[trim={0 100 0 50},clip,width=0.62\linewidth] {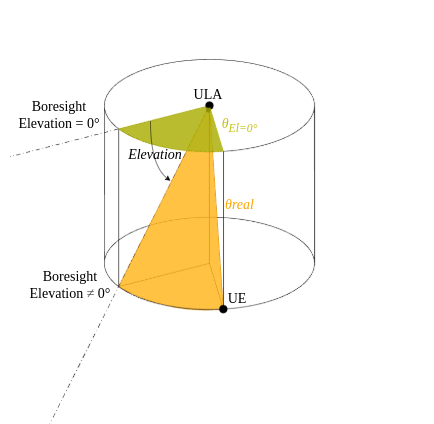}
    \caption{\small Geometric scenario of cylindrical correction. MUSIC naturally estimates the orange angle.}
    \label{fig: cylinder}
\end{figure}

An additional correction is applied to ensure that the positioning framework operates on the $XY$ plane, estimating the top-view azimuthal angle, green in Fig.~\ref{fig: cylinder}, rather than on the plane that intersects the \gls{ula} and the \gls{ue}, which corresponds to the raw \gls{aoa} (orange in Fig.~\ref{fig: cylinder}).

The correction formula can be expressed as:
\begin{equation}
\label{eq:correction_formula}
    \hat{\theta}_{XY} =  \arcsin{\biggl( \frac{d \sin\hat{\theta}}{\sqrt{d^2 - (\Delta z)^2}} \biggr) } , 
\end{equation}
where $d$ denotes the \gls{ula}-\gls{ue} distance, which can be obtained by integrating ranging information from a complementary positioning service, and $\Delta z$ represents the vertical offset between the \gls{ula} and the \gls{ue}. The height of the \gls{ula} is assumed to be known for each scenario, while the \gls{ue} height is fixed at $1.5$~m.


\begin{figure}[]
    \centering
    \includegraphics[width=\linewidth]{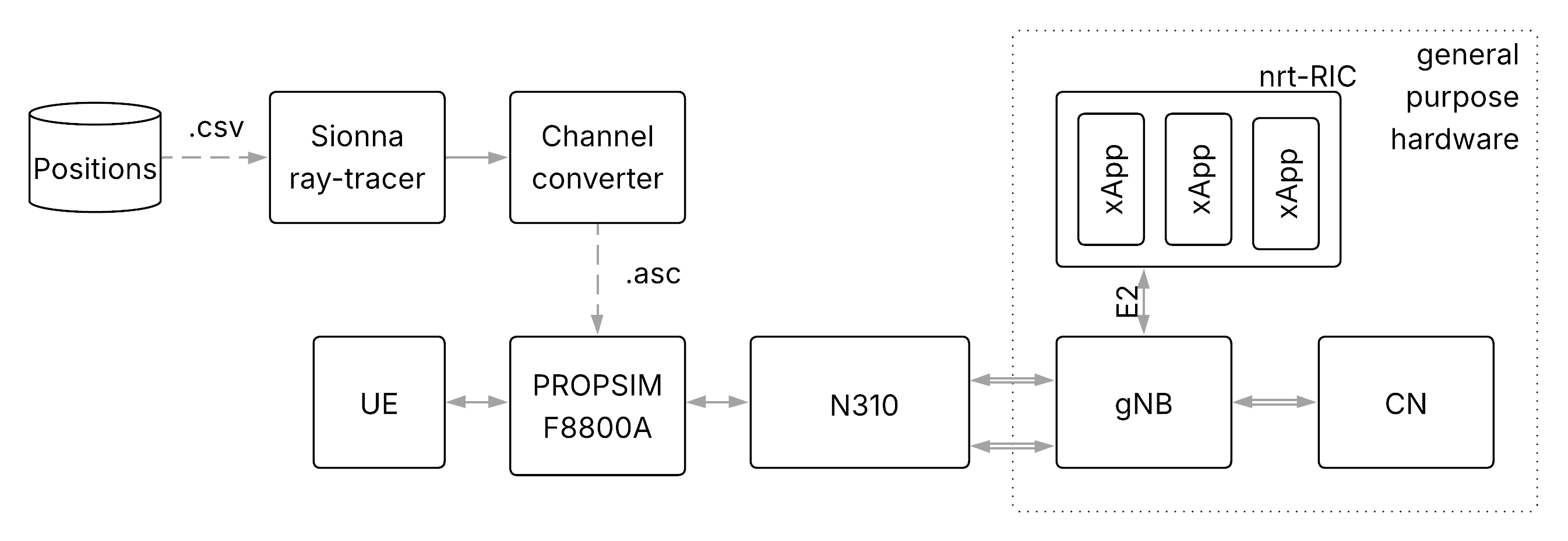}
    \caption{\small Experimental testbed: at the top, the Sionna ray tracer, which fetch the channels to the PROPSIM; to the left, the UE connected through a douplex connection to the O-RAN gNB.}
    \label{fig: siopsim}
\end{figure}

\section{Experimental Setup}%
\label{sec: setup}


To validate the proposed framework, we developed an experimental setup enabling a comprehensive evaluation of the integrated \gls{aoa} microservice across diverse propagation scenarios. The setup allows fine-tuning of channel characteristics to accurately reproduce realistic radio conditions.

Ray-tracing scenarios were designed in Sionna RT~\footnote{\url{https://developer.nvidia.com/sionna}} for channel characterization in specific environments. Each scenario includes a line-of-sight (LOS) component, along with specular and diffuse surface reflections. As the experiments were conducted at 3.95GHz, refraction effects were considered negligible. For each scenario, we modeled LOS as well as third- and fifth-order reflections to assess the impact of multipath propagation on \gls{aoa} estimation accuracy.

In all virtual scenarios, the \gls{gnb} was positioned at fixed coordinates $(x_{\textit{gNB}}, y_{\textit{gNB}}, z_{\textit{gNB}})$ for the entire duration of the experiments. It was equipped with a four-element isotropic antenna array arranged in a $1 \times 4$ planar configuration, with elements spaced by $\frac{1}{2}\lambda$. Conversely, the \gls{ue} employed a single vertically polarized isotropic transmitting element. The \gls{ue} was modeled in Sionna as a moving object following a trajectory specified in an external \texttt{.csv} file
, which provides its 
coordinates at discrete time steps $p_{\textit{UE}}(k) = k \cdot \Delta t$ for $k \in [0, 900]$, with $\Delta t = 0.1$~s. For each transmitter position, a path solver instance computed all \gls{ue}-\gls{gnb} propagation paths up to the specified interaction depth (i.e., reflection order). This process generated a set of multi-antenna, time-varying \glspl{cir} describing the propagation characteristics of a dynamic urban environment.

We then used Keysight PROPSIM F8800A channel emulator, a device capable of reproducing the exact propagation conditions of a target environment in real time, while remaining transparent to the connected devices. Channel emulation enables interaction with actual hardware and software as if they were operating in their native radio context, making it particularly valuable for testing scenarios that would otherwise be difficult, costly, or time-consuming to realize in practice. This setup allows us to combine the flexibility of simulation-based configuration with the realism of hardware-based validation. The \glspl{cir} generated in Sionna RT are converted into the PROPSIM-compatible \texttt{.asc} format using a dedicated channel converter developed for this framework.

For the validation phase, we integrated both commercial devices and devices based on open-source software, as depicted in Fig.~\ref{fig: siopsim}. The \gls{ue} is a \gls{cots} \gls{iot} device, a Sierra Wireless unit mounting a Qualcomm Snapdragon 888 \gls{5g} modem. 
Its transceiver is connected by coaxial cables to a port of the Keysight PROPSIM, which emulates the propagation channels. On the \gls{gnb} side, one emulator port is reserved for downlink traffic, while four ports are for uplink transmissions, corresponding to the four antenna elements of the \gls{ru}. The \gls{ru} is implemented using an Ettus Research N310 \gls{sdr} equipped with two daughterboards, each providing two transmit and two receive antenna ports. The \gls{sdr} streams digitized baseband data to the \gls{du} through a dedicated \acrlong{nic} using two 10~Gb/s \gls{sfp}+ optical links. The \gls{gnb} is deployed on a general-purpose server running the open-source \gls{oai} software stack and is connected to the Open5GS open-source 5G core network.

The programmability of the open RAN software stack enables the seamless integration of our dedicated solution within the \gls{oai} framework, allowing controlled experimentation with commercial devices and open-source network components under emulated yet realistic channel conditions.


\section{Experimental Results}%
\label{sec: results}


This section presents a systematic assessment of the proposed uplink \gls{aoa} estimation framework using the testbed presented in Sec.~\ref{sec: setup}.

The experiments were conducted across the aforementioned three scenarios designed to reproduce one of the most challenging geometries for wireless communication systems: urban canyons, which are known to degrade localization performance due to their multipath-rich propagation conditions. We evaluated the accuracy of the proposed framework over a total of 69,303~\gls{srs} signals exchanged between \gls{ue} and \gls{gnb}, $6\%$ of which correspond to \gls{nlos} conditions. Among all the signals exchanged, $31\%$ were in a zero multipath scenario, $34\%$ in a 3rd-order multipath environment, and $35\%$ in a 5th-order one.

\subsection{The impact of multipath}

\begin{figure}
\begin{subfigure}{\linewidth}
    \includegraphics[trim={0 2100 0 68},clip, width=\linewidth]{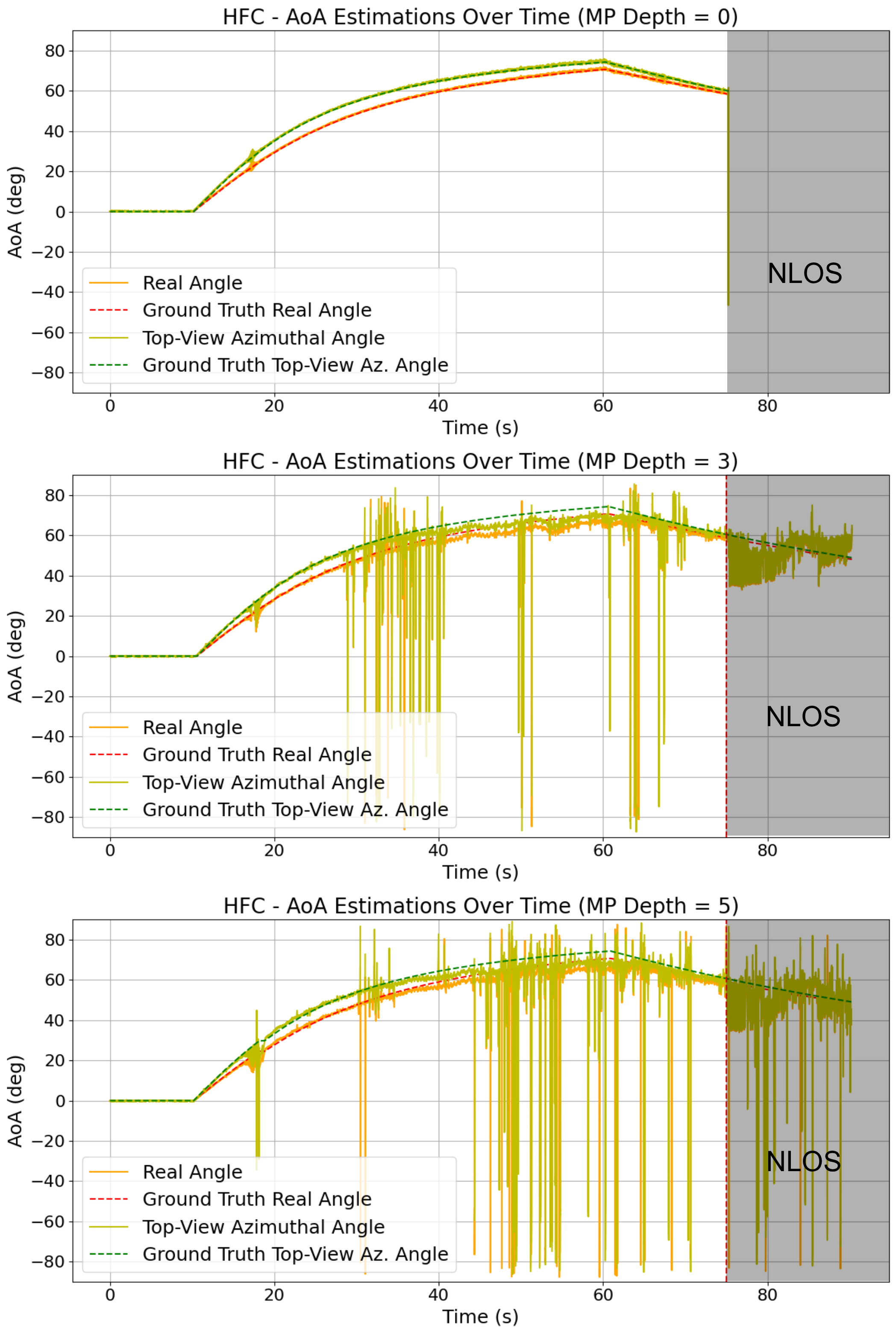} 
    \caption{No multipath scenario, LOS only.}
    \label{fig:hfc_0}
\end{subfigure}
\begin{subfigure}{\linewidth}
    \includegraphics[trim={0 18 0 2145},clip,width=\linewidth]{Content/figures/hfc_music_nodetach.png}
    \caption{Up to 5-th order reflections.}
    \label{fig:hfc_5}
\end{subfigure}
\caption{\small An example of AoA estimation with and without multipath for the same urban scenario. In orange the estimated $\hat{\theta}$, while in green, the corrected $\hat{\theta}_{XY}$.}
\label{fig:hfc_estimation}
\end{figure}

By progressively increasing the maximum reflection order, the channel is enriched with additional reflected components, allowing the analysis of the framework’s performance under increasingly complex and challenging propagation conditions.

Fig.~\ref{fig:hfc_estimation} illustrates a representative \gls{aoa} estimation trace obtained with (Fig.~\ref{fig:hfc_5}) and without (Fig.~\ref{fig:hfc_0}) multipath propagation. As the maximum reflection order increases, the number of secondary paths grows, leading to higher estimation variance and shifts in the MUSIC pseudospectrum. This behavior is evident from the numerous spikes observed in Fig.~\ref{fig:hfc_5}, which, however, can be effectively smoothed using a Kalman filter, as demonstrated in \cite{kalman_compare, kalman_turkish}.
This effect becomes particularly pronounced in \gls{nlos} conditions, highlighted in gray in Fig.~\ref{fig:hfc_estimation}. Although multipath propagation reduces the accuracy of \gls{aoa} estimation, it remains crucial for maintaining \gls{ue}–\gls{gnb} connectivity when the direct path is obstructed. As shown in Fig.~\ref{fig:hfc_0}, the \gls{ue} disconnects from the \gls{gnb} when both the line-of-sight component and sufficient power from reflected paths are absent, resulting in link failure.

\begin{figure}
    \centering
    \includegraphics[width=\linewidth]{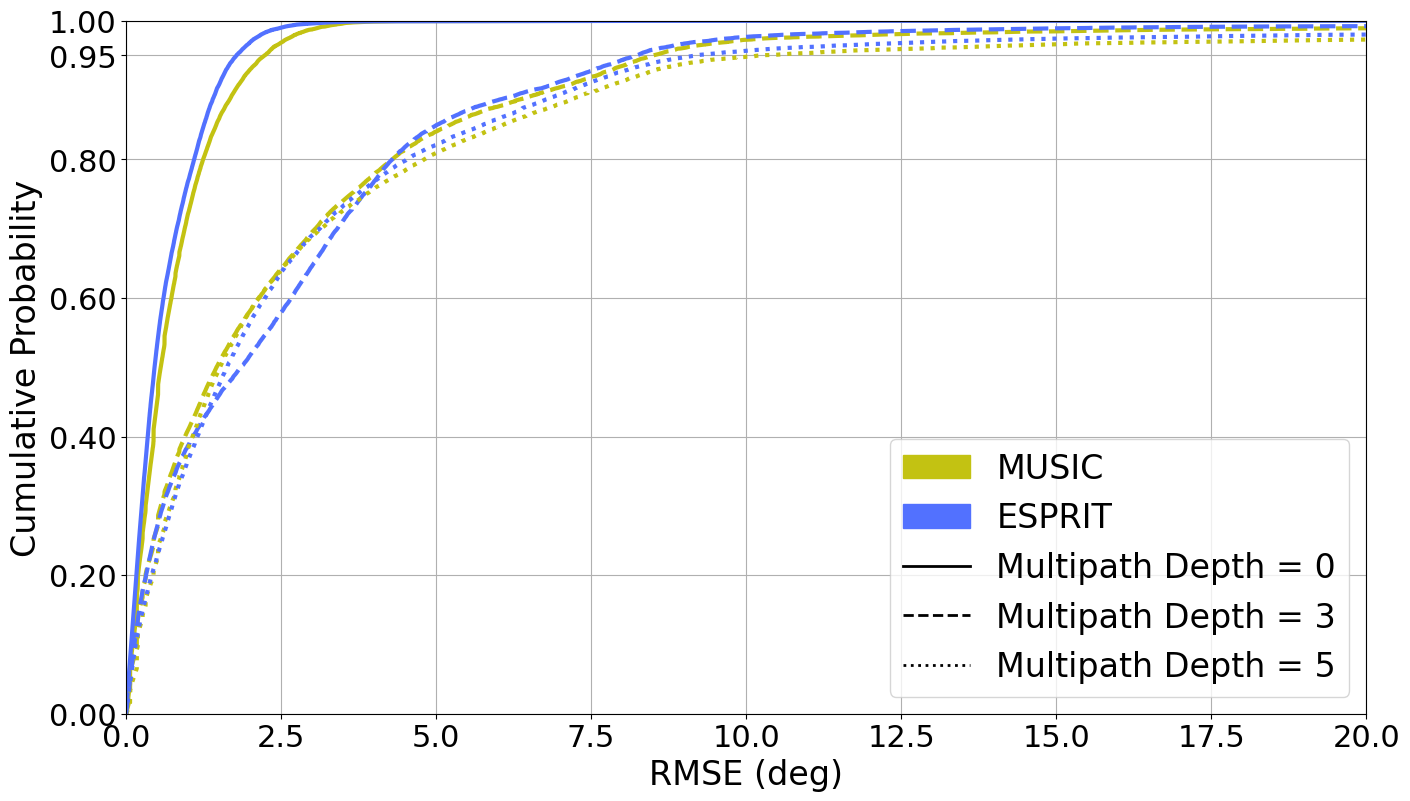}
    \caption{\small MUSIC vs ESPRIT in three different multipath depth scenarios, after cylindrical correction.}
    \label{fig:music_esprit}
\end{figure}

We further assess the impact of multipath using Fig.~\ref{fig:music_esprit}, which depicts the \gls{ecdf} of the \gls{rmse} for different reflection orders. The results confirm that in the absence of multipath (solid curves), the estimation error is tightly concentrated, with over 95\% of realizations exhibiting an \gls{rmse} below $2^\circ$. As the channel becomes progressively enriched with reflected components (dashed and dotted curves), the error distribution widens, revealing the degradation introduced by secondary propagation paths. Overall, the \gls{ecdf} analysis corroborates that although multipath is essential to sustain connectivity under \gls{nlos} conditions, it also constitutes the main source of estimation variability, particularly at higher reflection orders. Consequently, as expected, MUSIC and ESPRIT exhibit comparable accuracy under the same environmental conditions.

\subsection{The impact of SNR}



\begin{figure}
    \centering
    \includegraphics[width=\linewidth]{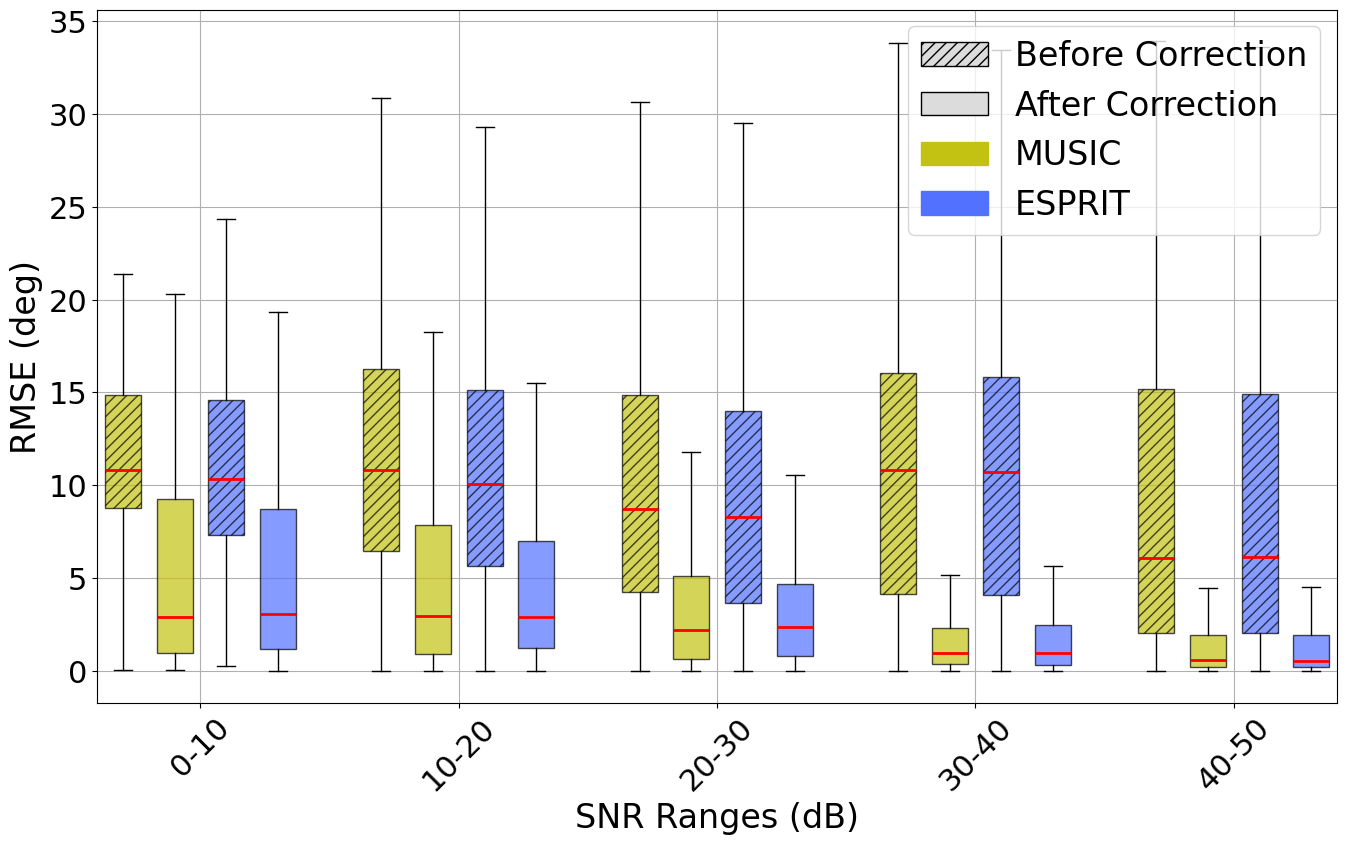}
    \caption{\small Distribution of RMSE Pre- vs Post- plane correction across different SNR ranges with MUSIC and ESPRIT AoA estimation.}
    \label{fig:music_esprit_box}
\end{figure}


The accuracy of the proposed framework improves as the \gls{snr} increases. Although this trend is expected, it is important to quantify the \gls{snr} range over which the \gls{aoa} service remains effective.
The boxplot in Fig.~\ref{fig:music_esprit_box} illustrates the distribution of \gls{aoa} \gls{rmse} across different \gls{snr} ranges. An improvement in bias-corrected estimation accuracy is observed as the \gls{snr} increases: for \gls{snr} values above 30~dB, over 95\% of the estimates exhibit an \gls{rmse} below $3^\circ$, while in the 0–10~dB range, achieving the same 95\% level requires tolerating errors up to $9^\circ$.
These results demonstrate that \gls{snr} is the dominant factor influencing \gls{aoa} estimation accuracy. However, even in multipath-rich environments, when the \gls{snr} is sufficiently high, MUSIC and ESPRIT algorithms consistently achieves sub-degree to few-degree precision. Conversely, under low-\gls{snr} conditions, noise degrades performance and increases variability; nevertheless, the algorithm continues to provide estimates that remain within practical accuracy bounds.

\section{Conclusions and Future Work}%
\label{sec: concl}


In this work, we presented a flexible and controllable testbed that, for the first time, enables systematic evaluation of \gls{aoa}-based positioning techniques under real yet fully repeatable channel conditions.
The proposed framework integrates NVIDIA Sionna RT with the Keysight PROPSIM channel emulator and includes a novel phase calibration procedure for the USRP~N310. By combining simulation flexibility with hardware realism, the testbed allows the experimental validation of \gls{aoa} estimation algorithms in realistic propagation environments, providing results of direct relevance for next-generation \gls{5g} deployments.

The proposed opportunistic calibration method proved effective but still requires manual initialization. Future work will focus on automating this process through self-calibration mechanisms that exploit known pilot signals transmitted and analyzed by the \gls{gnb} itself. Such automation would reduce operator intervention and ensure long-term phase stability, enabling fully autonomous operation of the testbed.

Finally, the \gls{aoa} estimation algorithm has been implemented for deployment as an \gls{oran} xApp, enabling seamless integration with complementary \gls{toa}-based frameworks. This makes the proposed platform a practical and extensible foundation for network-native \gls{5g} localization services.

%

\bibliographystyle{IEEEtran}
\bibliography{only}

@inproceedings{bernazzoli2025robust,
    author = {Bernazzoli, Viola and {et al.}},
    title="{Robust Uplink Ranging in 5G Networks: An Integrated O-RAN Approach}",
    booktitle={IEEE MASS}, 
    year={2025}
}

@phdthesis{invariance_shift,
    author = "Martin Haardt",
    title = "Efficient one-, two-, and multidimensional high resolution array signal processing",
    school = "Technische Universitat Munchen",
    year = 1997
}

@ARTICLE{music,
       author = {{Schmidt}, R.~O.},
        title = "{Multiple emitter location and signal parameter estimation}",
      journal = {IEEE Transactions on Antennas and Propagation},
         year = 1986}

@manual{n310_user,
    authorgroup={Ettus Research},
    url = {https://files.ettus.com/manual/page\_usrp\_n3xx.html},
    urldate={09/22/2025},
    title = {USRP Hardware Driver and USRP Manual},
    year={2025},
    version={4.9.0.0}
}

@article{joint_LTE,
  author={Shamaei, Kimia and Kassas, Zaher M.},
  journal={IEEE TSP},
  title={A Joint TOA and DOA Acquisition and Tracking Approach for Positioning With LTE Signals}, 
  year={2021}}

@article{sun_comparative_single_gnb,
AUTHOR = {Sun, Bo and {et al.}},
TITLE = {A Comparative Study of 3D UE Positioning in 5G New Radio with a Single Station},
JOURNAL = {Sensors},
YEAR = {2021}}

@techspecs{38211,
    author = {{3GPP}},
    title = {5G; NR; Physical channels and modulation, (TS {38.211} v18.6.0)},
    year = 2025
}

@paper{5g_aoa_techniques_comparizon,
      title={Angle of Arrival Estimation Using SRS in 5G NR Uplink Scenarios}, 
      author={Spanos, Thodoris and {et al.}},
    url={https://arxiv.org/abs/2411.16501},
      year={2024}}

@article{xhafa_comparizon,
  author={Xhafa, Alda and {et al.}},
  journal={IEEE Transactions on Instrumentation and Measurement}, 
  title={Experimental Investigation of AoA Estimation and Antenna Calibration With 5G NR SignalsUsing USRP Devices}, 
  year={2025}}

@articel{jade,
  author={Vanderveen, M.C and {et al.}},
  journal={IEEE Communications Letters}, 
  title={Joint angle and delay estimation (JADE) for multipath signals arriving at an antenna array}, 
  year={1997}}

@inproceedings{joint_li,
  author={Li, Yiwen and {et al.}},
  booktitle={IEEE VTC}, 
  title={5G Communication Signal Based Localization with a Single Base Station}, 
  year={2020}}

@article{iot_survey,
  author={Akpakwu, Godfrey Anuga and {et al.}},
  journal={IEEE Access}, 
  title={A Survey on 5G Networks for the Internet of Things: Communication Technologies and Challenges}, 
  year={2018}}

@book{gnss_perform,
  author={Kaplan, Elliott and {et al.}},
  title={Understanding GPS/GNSS: Principles and Applications, Third Edition},
  year={2017}}

@book{aoa_bible,
  title={Introduction to Direction-of-arrival Estimation},
  author={Chen, Z. and Gokeda, G. and Yu, Y.},
  isbn={9781596930902},
  lccn={2010278240},
publisher={Artech House},
year={2010},}

@article{oran_polese,
  author={Polese, Michele and {et al.}}, 
  journal={IEEE Commun. Surv. Tutor.}, 
  title={Understanding {O-RAN}: Architecture, Interfaces, Algorithms, Security, and Research Challenges}, 
  year={2023}}

@article{kalman_compare,
author = {Venu, M.},
year = {2015},
title = {A Comparative Study of DOA Estimation Algorithms with Application to Tracking Using Kalman Filter},
journal = {Signal \& Image Processing: An International Journal},}

@inproceedings{kalman_turkish,
  author={Yildiz, Umut and Coşkun, Ahmet Faruk},
  booktitle={Signal Processing and Communications Applications Conference}, 
  title={Application of Kalman Filtering Approach on Signal Angle-of-Arrival Estimates}, 
  year={2023}}

@misc{sionna,
      title={Sionna: An Open-Source Library for Next-Generation Physical Layer Research}, 
      author={Hoydis, Jakob and {et al.}},
      year={2023},
    url={https://arxiv.org/abs/2203.11854}}

\end{document}